\documentclass{pasj00}

\newcommand{\Tin}{T_{\rm in}^{\rm obs}}
\newcommand{\Tinm}{T_{\rm in}^{\rm mod}}
\newcommand{\Rin}{R_{\rm in}^{\rm obs}}
\newcommand{\Rinm}{R_{\rm in}^{\rm mod}}
\newcommand{\rg}{r_{\rm g}}
\newcommand{\rS}{r_{\rm g}}

\begin{document}
\SetRunningHead{K. Watarai and S. Mineshige}
{Slim Disk: Viscosity Prescriptions and Observational Implications}
\Received{2001/2/28}
\Accepted{2001/8/14}

\title{Slim Disk: Viscosity Prescriptions and Observational Implications}

\author{Ken-ya \textsc{Watarai}}
\affil{Department of Astronomy, Graduate School of Science, 
Kyoto University, Sakyo-ku, Kyoto 606-8502}
\email{watarai@kusastro.kyoto-u.ac.jp}

\author{Shin \textsc{Mineshige}}%
\affil{Yukawa Institute for Theoretical Physics, 
Kyoto University, Sakyo-ku, Kyoto 606-8502}
\email{minesige@yukawa.kyoto-u.ac.jp}
%

\KeyWords{accretion: accretion disks, black holes---stars: X-rays} 

\maketitle

\begin{abstract}
We examine the effects of the different viscosity prescriptions and
the magnitude of the viscosity parameter, $\alpha$, on the structure of the slim disk, 
and discuss the observational implications on accretion-flow structure into a stellar-mass black hole.
In contrast with a standard disk, in which the 
``$\alpha$'' value does not affect significantly the local flux,
radiation from the slim disk is influenced by
the $\alpha$ value.
For the range of $\alpha = 10^{-2} \sim 10^{0}$ we calculate the disk 
spectra and from spectral fitting we derive $\Tin$, maximum temperature of
 the disk, $\Rin$, the size of the region emitting blackbody radiation
 with $\Tin$, and $p\equiv -{\rm dln} T_{\rm eff}/{\rm dln}~r$, the
 slope of the effective temperature distribution. 
It was founded that the estimated $\Tin$ slightly increases
as $\alpha$ increases.
This is because the larger the magnitude of viscosity is,
the larger becomes the accretion velocity and, hence,
the more enhanced becomes advective energy transport,
which means less efficient radiative cooling and 
thus higher temperatures.

Furthermore we check different viscosity prescriptions
with the form of the viscous stress tensor of 
$t_{r \varphi} = -\alpha \beta^{\mu}p_{\rm total}$,
where $\beta$ is the ratio of gas pressure
($p_{\rm gas}$) to total pressure, $p_{\rm total}$
($= p_{\rm gas}+p_{\rm rad}$), and $\mu$ is a parameter ($0 \le \mu \le 1$).
 For $\mu=0$ we have previously found that as luminosity approaches
 the Eddington , $L_{\rm E}$,$\Rin$ decreases below 3$\rg$ 
($3r_{\rm g}$ corresponds to the radius of the marginally stable circular orbit, 
$r_{\rm ms}$, with $r_{\rm g}$ being Schwarzschild radius) and the
 effective temperature profile becomes flatter, 
$T_{\rm eff} \propto r^{-1/2}$. 
Such a slim-disk nature does not appear when $\mu$ is large, $\mu \sim 0.5$, even at the Eddington luminosity. 
Hence, the temperature of the innermost region of
the disk sensitively depends on the $\mu$ value. 
We can rule out the case with large $\mu~(\sim 0.5)$, since
it will not be able to produce a drop in $\Rin$
with an increase in luminosity as was observed in an ultraluminous X-ray
 source, IC~342, source 1.

\end{abstract}

\section{Introduction}


The accretion disk model by Shakura and Sunyaev (1973, hereafter, SSD)
has been regarded as the ``standard'' model for the low mass X-ray
Binaries (LMXBs) and galactic black-hole candidates (GBHCs)
in their soft state.
Their analytical solutions are self-consistent under some specific
assumptions.  However, when its luminosity $L$ approaches
the Eddington luminosity, $L_{\rm E}$, some assumptions break down and
thus the standard model cannot adequately describe the nature 
of such high luminosity objects.
The critical luminosity is $L \sim 0.3 L_{\rm E}$ (Abramowicz et al. 1988, Laor, Netzer 1989).
This fact is theoretically established but 
not widely known and still most of investigations have
been conducted by simply applying the canonical SSD even to
near-Eddington luminosity systems without any comments.

Recent observations have detected several kinds of exotic
super-Eddington sources, whose luminosity seemingly exceeds the Eddington luminosity.
The properties of their energy source
cannot be described by the SSD theory, and hence
its origin is still an open astrophysical question.
The nature of super-Eddington sources has been discussed
in relation to their X-ray properties since late 1980's 
(Fabbiano, 1989). 
For instance, Ultra-Luminous Compact X-ray Sources (hereafter, ULXs) in
nearby spiral galaxies are very mysterious objects.
Their luminosities amount to around $10^{39-40} {\rm erg~s^{-1}}$, which
means the mass ($M$) of black holes, if they exist, should exceed 10--100
$M_{\odot}$. 
However, the MCD (multi-color disk, Mitsuda et al. 1984)
fitting derives relatively high temperatures, $\Tin \simeq 1.0-2.0$ keV
(Mizuno et al. 1999, Makishima et al. 2000), compared with the
theoretical expectations based on the SSD model, which is significantly
below 1.0 keV for $M \gtrsim 100 M_{\odot}$.
Here, $\Tin$ is the color temperature derived from the observed
spectrum (for the exact definition, see Section 3.3.). 
In our Galaxy, there exists a famous galactic microquasar, GRS 1915+105,
which shows the complicated time/spectral variation.
Its luminosity is near the Eddington luminosity in its high state $\gtrsim 
a few \times 10^{39} {\rm erg~s^{-1}}$ (Belloni et al. 2000). 
These objects can never be explained in the framework of the SSD theory.

On the theoretical side,
there has been growing recognition
that the super-critical accretion disk model, so-called ``slim disk''
(Abramowicz et al. 1988) may be a correct model to describe these high
luminosity objects. Its significance has been re-recognized
in connection with luminous AGNs
(Szuszkiewicz et al. 1996, Wang et al. 1999),
narrow-line Seyfert1 galaxies (NLS1s; Mineshige et al. 2000),
and GBHCs shining near the Eddington luminosity (Fukue 2000,
Watarai et al. 2000), since the disk is stable due to the advective
entropy transport, even at the super-critical accretion rates.
The ULXs are discussed along this line in the first time by Watarai et
al. (2001) and it has been shown to explain the basic observed tendency,
although more detailed analysis is needed (Watarai in preparation).

The local stability against thermal perturbations can 
be discussed in terms of the thermal equilibrium curve, 
so called ``S-shaped" curve (Taam, Lin 1986, Abramowicz et al. 1995). 
The local stability properties for the stationary slim disk models with different viscosity prescriptions were discussed by Szuszkiewicz (1990), who found that when ``$\mu$" increases the thermally unstable 
region in the disk becomes smaller.
Here, we prescribe viscous-shear tensor,
the different form of shear stress tensor to be, 
$t_{r \varphi} = -\alpha p^{\mu}_{\rm gas} p^{1-\mu}_{\rm total}$, 
($p_{\rm gas}$ is the gas pressure, 
and $p_{\rm total}$ is the total pressure), 
where $\mu$ is the parameter ($0\le \mu \le1$). 

The global stability analysis for slim disk model 
has been carried out by Honma et al. (1991) 
who found that the disk was stabilized as the viscosity parameter 
``$\mu$" increased. 
Szuszkiewicz and Miller (1997) confirmed this finding, showing  that the models are stable not only when they are locally stable but also when the local analysis predicts an unstable region with radial dimension smaller than the shortest wavelength of the unstable modes.
In the most recent study by Zampieri et al. (2001) 
have calculated the spectra of a slim disk model 
along one full thermal limit cycle. 
However, there is little research that carried out direct comparison with 
X-ray observations.

%
Our concern in the present study is how the viscosity prescription and its strength act on
the structure of accretion disks, and how they manifest themselves in
the fitting parameters ($\Tin$, $\Rin$) derived from the observational spectra in soft X-ray band. 
If $\Tin$ and $\Rin$ can be accurately determined from the observation,
conversely, the functional form of disk viscosity might be able to constrain.
A comprehensive understanding of the structure and spectrum of slim disks is an outstanding issue.

In the next section we introduce the basic equations and explain our calculation method.
In section 3, the results of our calculation are presented for
a variety of viscosity parameters ($\alpha, \mu$).
The observational features of the slim disk are summarized by using the
fitting parameters, $\Tin, \Rin$, and $p$ in section 4.
Final section is devoted to discussion and concluding remarks.

\section{Basic Equations}

Basic equations we use here are all the same as those in 
the previous paper (Watarai et al. 2000) 
except for the viscosity prescription (see Kato et al. 1998, Chapter 8
for the detailed description). 
We solved the height-integrated equations (H$\bar{\rm o}$shi, 1977) 
in the radial direction on the basis of pseudo-Newtonian potential 
(Paczy\'{n}sky, Wiita, 1980), and
neglected self-gravity. 
The pseudo-Newtonian potential is not an excellent approximation around the event horizon $\sim \rg$. 
Thus, caution is needed when the derived $\Rin$ is small, $\Rin \lesssim
\rg$. Other general relativistic effects needs also to take account, then. Otherwise, our approximation does not introduce serious errors.

Honma et al. (1991) were the first to carry out the calculation above 
circumstances (including the effect of viscosity prescriptions of the
slim disk). 
The main difference of our calculation and theirs is that 
they truncated the disk inner edge at 2.7 $r_{\rm g}$, 
while we integrate the disk down to the real vicinity 
of the event horizon at $r_{\rm g}$. 

The momentum equation in the radial direction is  
\begin{equation}
\label{r-mom}
   v_r{dv_r\over dr}+{1 \over \Sigma}{dW\over dr}
   ={\ell^2-\ell_{\rm K}^2\over r^3}
   -{W\over \Sigma}{d{\rm ln}~\Omega_{\rm K}\over dr},   
\end{equation} 
where $v_{r}$ is the radial velocity of gas, 
$\Sigma=2 I_{\rm 3} \rho_0 H$ is surface density,  
and $W \equiv \int p_{\rm total} dz = 2 I_{4} p_0 H$ is pressure
integral, $H$ is disk scale height, respectively. 
Here, the pressure, $p_0$, and the density, $\rho_0$, are related to 
 each other by the relation $p_0 \propto \rho_0^{1+1/N}$ 
(the subscript `0' indicates the quantities on the equatorial plane). 
We fixed the polytropic index to be $N=3$, for which the numerical
constants are $I_3$=16/35 and $I_4$=128/315 (H$\bar{\rm o}$shi, 1977).

The specific angular momentum and the Keplerian angular momentum are 
defined by $\ell$ ($=rv_\varphi$) and $\ell_{\rm K}=r^2\Omega_{\rm K}$,
respectively, where $\Omega_{\rm K}$ is the Keplerian angular frequency. 
The angular momentum conservation is 
\begin{equation}
\label{ang-mom}
    {\dot M}(\ell-\ell_{\rm in})=-2\pi r^2T_{r\varphi},  
\end{equation}
where $\dot{M}$ is mass accretion rate, 
\begin{equation}
   {\dot M} = -2\pi r v_r \Sigma  = {\rm constant.},    
\end{equation}
from continuity equation and 
$T_{r\varphi}$ is the height integrated viscous stress tensor, 
$T_{r\varphi}\equiv \int t_{r\varphi} dz$. 

Hydrostatic balance in the vertical direction leads 
\begin{equation}
\label{h-balance}
    H^2\Omega_{\rm K}^2= (2N+3)\frac{W}{\Sigma} = c_{\rm s}^{2},
\end{equation}
where $c_{\rm s}$ is the sound speed. 

The equation of state is 
\begin{equation}
\label{pressure}
   p_{\rm total}= p_{\rm gas} + p_{\rm rad} = 
   \frac{R}{\bar{\mu}} \rho_0 T_0 + \frac{a}{3}T_{0}^4,    
\end{equation}
where $R,~\bar{\mu}$, and $a$ are the gas constant, 
the mean molecular weight, and the radiation constant respectively. 
As to the viscosity prescription 
we rewrite it in terms of the effective $\alpha$;
\begin{equation}
\label{alp1}
    T_{r \varphi} = -\alpha_{\rm eff} W ~~ 
{\rm with~}
     \alpha_{\rm eff} \equiv \alpha \beta^{\mu} 
            = \alpha (p_{\rm gas}/p_{\rm total})^\mu. 
\end{equation}
This form is convenient to compare with the cases with the usual
viscosity prescription (i.e., $\mu = 0$).
If we take an extreme case of $\mu \to 0$ limit, 
the stress tensor is proportional to total pressure; 
i.e., the same as that of the SSD prescription. 
In the large $\mu$ limit $(\mu \to 1)$, on the other hand, 
the shear stress tensor depends solely on gas pressure.
This situation may realize when there grows turbulence generated by chaotic magnetic fields 
(Lightman, Eardley 1974; Sakimoto, Coroniti 1981). 

The energy equation involves the viscous heating, 
radiative cooling and advective cooling terms; namely,
\begin{equation}
\label{energy}
  Q_{\rm vis}^{+}=Q_{\rm rad}^{-}+Q_{\rm adv}^{-}, 
\end{equation}
where each term is explicitly written as:
\begin{eqnarray}
\label{q_vis}
    {Q_{\rm vis}^+} &=& r T_{r\varphi}\frac{d\Omega}{dr} 
= -r \alpha_{\rm eff} W \frac{d\Omega}{dr}, \\  
\label{q_rad}
    {Q_{\rm rad}^-} &=& \frac{8 a c T_{0}^4}{3 \bar{\kappa} \rho_{0} H}, \\  
{\rm and} \nonumber  \\
\label{q_adv}
    {Q_{\rm adv}^-} &=& \frac{I_4}{I_3} v_r\Sigma T_0 \frac{ds_0}{dr}. 
\end{eqnarray}
Here, $s_0$ is the entropy on the equatorial plane,
$\bar{\kappa}$ is the Rosseland-mean opacity,   
$\bar{\kappa} = \kappa_{\rm es}+\kappa_{\rm ff} = 0.40 + 0.64 \times 
   10^{23} {\bar{\rho}}~{\bar{T}}^{-7/2} {\rm g^{-1} cm^{2}} $
with $\bar{\rho}$ and $\bar{T}$ being averaged density and 
temperature, respectively,
$(\bar{\rho}=16/35 \rho_0$ and $\bar{T}=2/3 T_0)$ and other symbols have
their usual meanings.
We solved the above set of equations (\ref{ang-mom})--(\ref{energy}). 
The solution is calculated 
from the outer edge of the disk ($r_{\rm out}=10^4 r_{\rm g}$) 
to the vicinity of the central black hole through the transonic point. 
We adjust $l_{\rm in}$, specific angular momentum
which is finally swallowed by the black hole,
so as to satisfy the regularity condition at the transonic radius.

For evaluating the effective temperature, we assume that 
all the radiative energy is emitted with blackbody as a first approximation. 
 Namely, using the Stefan-Boltzmann's law, 
$Q_{\rm rad}^- = 2 F = 2 \sigma T_{\rm eff}^4$, 
we define the local effective temperature to be 
$T_{\rm eff} = (Q_{\rm rad}/{2 \sigma})^{0.25}.$
As is widely known, however, the effect of Compton scattering is
substantial in electron scattering dominated layers (Czerny, Elvis
1987; Ross et al. 1992; Shimura, Takahara 1995; Wang et al. 1999). 
We include this effect by introducing a spectral hardening factor (see
Section. 3.3.). 
Further, photon trapping effect 
(e.g., Mineshige et al. 2000) needs to be considered. 
However, we ignore these effects in order to understanding the basic
tendency of the spectrum.

Finally, we define the normalized parameter 
$\dot{m} = \dot{M}/\dot{M}_{\rm crit}$, with 
$\dot{M}_{\rm crit} = L_{\rm E}/c^2 
= 1.3 \times 10^{17} (M/M_\odot) {\rm g~s^{-1}}$.
This definition may differ in different literatures. 
In our notation, disk luminosity corresponds to the Eddington luminosity,
when the accretion rate is $\dot{m}=16$
(provided that the pseudo-Newtonian potential is used).
We vary the normalized parameters, $m \equiv M/M_{\odot}$, 
$\dot{m}$, $\alpha$, and $\mu$, and calculate the structure of 
optically thick accretion disk for a wide range of parameters.

\section{The effects of varying viscosity prescriptions}

\subsection{The $\alpha$-Dependence}

We first examine the effects of changing the $\alpha$ parameter 
within the SSD prescription, $t_{r \varphi}=-\alpha p_{\rm total}$, 
i.e., $\mu=0$.
It is well known that $\alpha$ does 
not affect significantly the spectrum in the standard disk theory, 
since total radiation output rate is balanced with the energy 
generation rate via viscous processes and does not explicitly depend on density nor velocity. 
In other words, effective temperatures can be nearly uniquely determined
by black-hole mass and mass-accretion rate as functions of radius.
In a high accretion-rate system, in contrast, $\alpha$ dependence appears,
since radiative cooling is no longer balanced solely with viscous
heating so that the disk inner temperature can change even for the same
$M$ and $\dot{M}$.

\begin{figure}
  \begin{center}
  \begin{tabular}{cc}
    \FigureFile(80mm,80mm){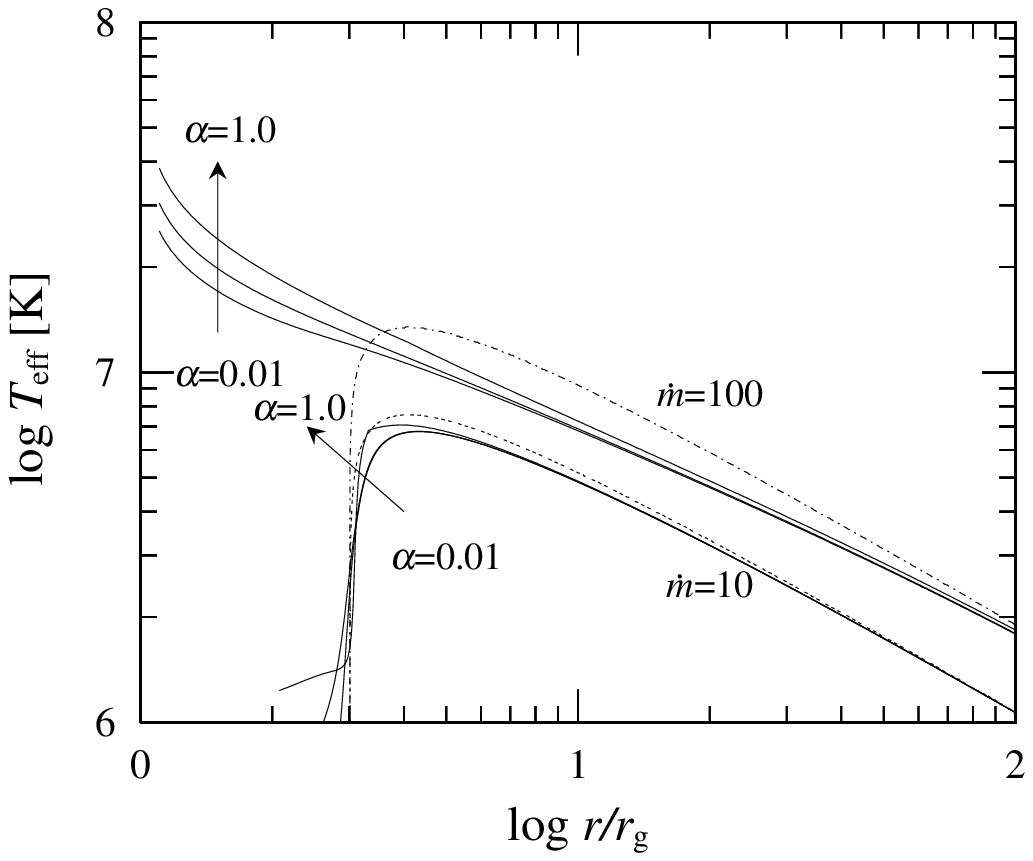} \\ 
    \FigureFile(80mm,80mm){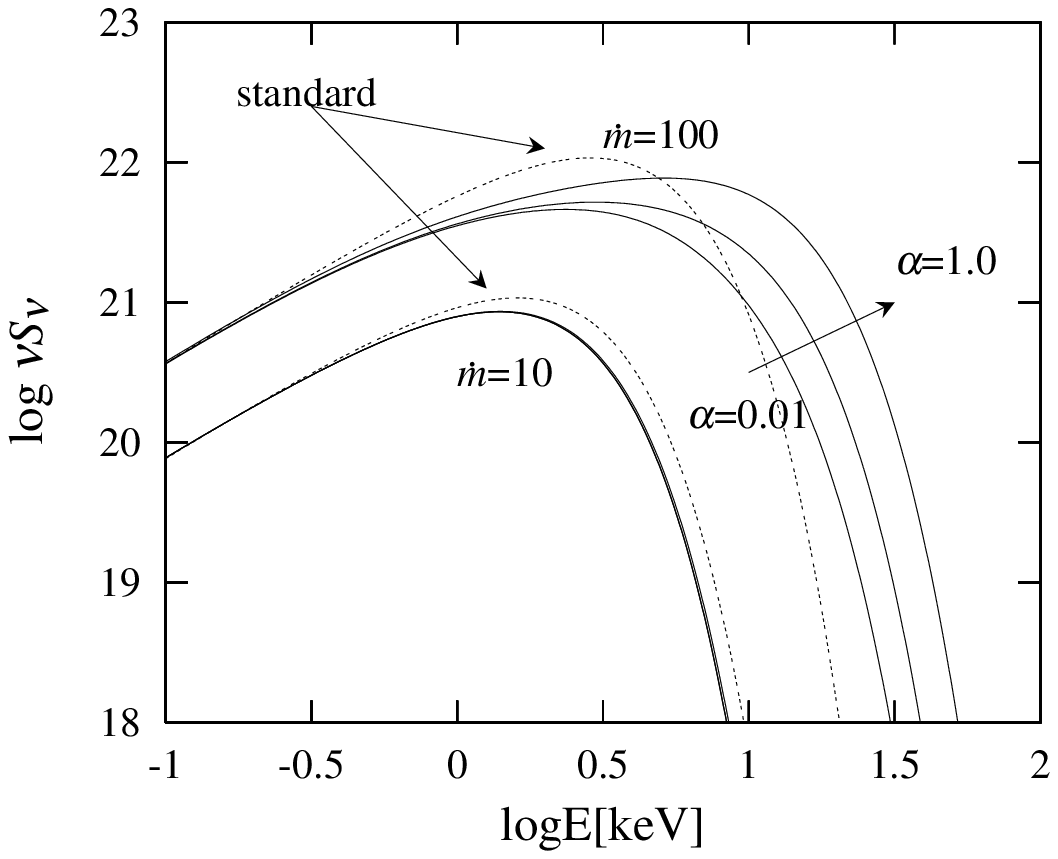}    
  \end{tabular}
  \end{center}
  \caption{The effective temperature distributions 
and the emergent spectra for different $\alpha$ parameters
 ($\alpha$=0.01, 0.1, and 1.0). 
Solid lines represent the calculated models (for $\dot{m}=10$ and $100$). 
Dashed lines are the results based on the standard-disk relation (Shakura, Sunyaev 1973).
For a small accretion rate, $\dot{m}=10$, 
the spectrum does not largely depend on $\alpha$ parameter, where as  
at a high accretion rate, $\dot{m}=100$,  
$\alpha$ dependence appears in the temperature profile in the inner region. 
Likewise, 
the spectral changes are not appreciable in the case of $\dot{m}=10$ while in $\dot{m}=100$ case, 
the spectrum has $\alpha$ dependence reflecting the $\alpha$ dependence
 of the temperature distribution.
Black-hole mass is fixed to be 10$M_{\odot}$.}
\label{tesp_alpha}
\end{figure}

\begin{figure}
  \begin{center}
    \FigureFile(80mm,80mm){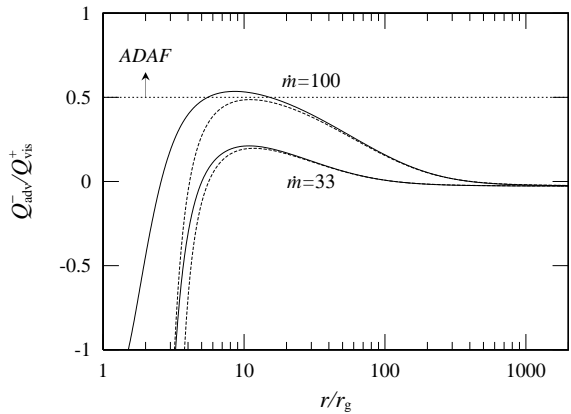} 
  \end{center}
  \caption{
The ratio of advective energy transport to the viscous
 heating rate for different accretion rate ($\dot{m}$=33 and 100). 
For all the cases, the viscous heating rate $Q_{\rm vis}^+$ is balanced 
with the radiative cooling $Q_{\rm rad}^-$ at larger radii, thus
 $|Q_{\rm adv}^-| \ll Q_{\rm vis}^+$. 
At higher accretion rates,
the advective cooling $Q_{\rm adv}^-$ is substantial at inner radii, 
but still, the ratio is mostly less than 0.5;that is, the flow is not advection dominated. 
Black-hole mass is fixed to be $M=10M_{\odot}$, and we set $\alpha=0.1$
 (the solid lines) and $0.01$ (the dotted lines).
}
\label{fig:energy}
\end{figure}

Figure \ref{tesp_alpha} shows the temperature profiles and spectra
for different $\alpha$ and $\dot{m}$. 
In the temperature profiles, we found $T \propto r^{-3/4}$ at low
$\dot{m} \lesssim 10$ (called as the SSD regime) 
and $T \propto r^{-1/2}$ at large $\dot{m} \gtrsim 100$ 
(called as the slim-disk regime, see Watarai, Fukue, 1999, Watarai et
al. 2000; Wang et al. 1999). 
For instance, when $\alpha$ increases by one order of magnitude,
the temperature in the vicinity of the disk inner region only slightly
rises by a factor of $\sim 2$,
and X-ray radiation spectra get a bit harder.

Next, we check the relative importance of advective energy transport in
the energy equation. 
Figure \ref{fig:energy} plots the ratio, $Q_{\rm adv}^-/Q_{\rm vis}^+$, for different $\alpha$ and $\dot{m}$.
When $\alpha$ increases, this ratio also increases; 
that is, the large $\alpha$ is, the more effectively
advection cooling works.
We note, however, that even for $\dot{m}=100$, 
still disk is not entirely advection dominated 
(the maximum ratio is $\sim 0.5$).
Thus, it is not appropriate to call the slim disk the optically thick
ADAF (see also Abramowicz et al. 1988).

Figure 1 explicitly demonstrates that the disk inner edge decreases as
$\dot{m}$ increases. 
Why, then, can  the inner edge of the disk be smaller than the radius of the marginally stable circular orbit, $r_{\rm ms} \sim 3\rg$? 
One of the reasons is a decrease of the transonic radius of the flow,
$r_{\rm S}$, ($r_{\rm S} \lesssim  r_{\rm ms}$) with increase of
accretion rate; i.e. $r_{\rm S}/\rg=$3.00, 2.84 and 2.69 for $\dot{m}\le
10, 33$ and 100, respectively.
We obtained $r_{\rm S}$ for several $\dot{m}$ and confirmed that the
tendency of $r_{\rm S}$ is consistent with that of Abramowicz et al. (1988).  
In the case of super-critical accretion, moreover, a pressure gradient
force is important as in the case of a thick torus (Abramowicz et al. 1978). 
For small $\alpha$ ($<$ 0.1) accretion is, in fact, driven by pressure
gradient. 
Then, the pressure gradient force tend to push material near the inner edge
of the disk inward to smaller radii, even if the material has larger
angular momentum than the equilibrium value, for which centrifugal force
is balanced with gravity force.
Finally, even when the flow is unstable for circular motion, the growth
time is on the order of free-fall accretion time, while in the meantime mass is continuously
supplied from outside with large (comparable to sound) velocity at
large $\dot{m}$. 
There always exists substantial material inside $r_{\rm ms}$, emitting radiation. 

For large $\alpha$ value ($\sim 1.0$) the accretion is driven by the viscosity, however, such a flow can be effectively optically thin.
It was pointed out by Beloborodov (1998) that
in the case of very large $\alpha\sim 1.0$, 
the inner region of the disk may be overheated due to decreasing the 
effective optical depth, but strong convective
motion will reduce thermalization timescale, 
which may avoid overheated problem for moderately large $\alpha \sim 0.1$ (Mineshige et al. 2000).

\subsection{The $\mu$ Dependence}

In this subsection, we discuss the $\mu$ dependence 
of the slim disk structure.  In equation (\ref{alp1}), 
we fix $\alpha$ (=0.1) and $M~(=10M_\odot)$ 
but change the value of $\mu$ (=0, 0.1, 0.4, and 0.49). 
If we take the limit of $\mu$ approaching unity; that is 
$t_{r \varphi}=-\alpha p_{\rm gas}$, 
and then the disk will be completely stabilized, since
then temperature dependence of the heating rate ($\propto t_{r \varphi}$)
is now less than that of the cooling term, a condition
for thermal stability (see Kato et al. 1998, Chap. 4).

\begin{figure}
  \begin{center}
  \begin{tabular}{cc}
    \FigureFile(80mm,80mm){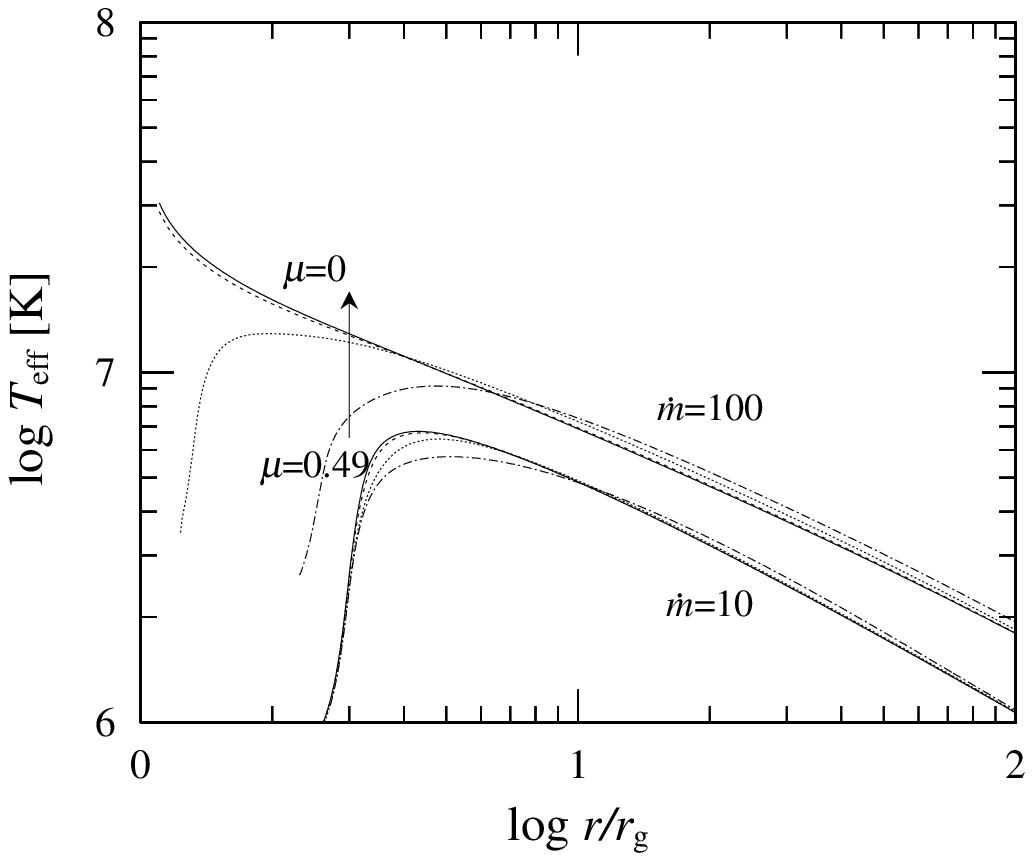} \\ 
    \FigureFile(80mm,80mm){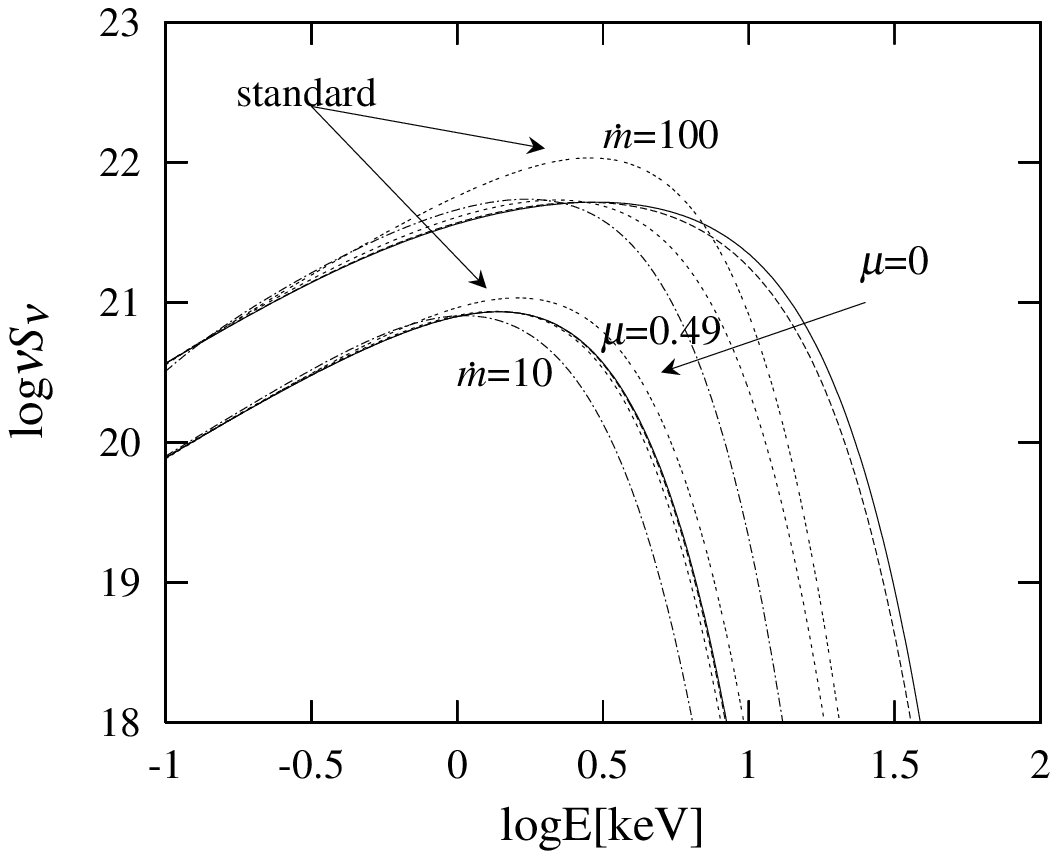}    
  \end{tabular}
  \end{center}
  \caption{The same as figure 1 but different $\mu$ parameters
($\mu$ = 0, 0.1, 0.4, and 0.49).}
  \label{temu}
\end{figure}

Figure \ref{temu} display the temperature profiles and 
the spectra for different $\mu$ values.
When $\dot{m}=10$, the temperature does not show $\mu$ dependence. 
In the $\dot{m}=100$ case, in contrast, 
the temperature appreciably decreases around transonic region ($ <
3r_{\rm g}$) as $\mu$ decreases, keeping the same temperature slope, $T
\propto r^{-1/2}$, at large radii, $r \gg 3\rg$. 

The basic interpretations of these results are that 
a large $\mu$ tends to decrease $\alpha_{\rm eff}$ at smaller radii, 
thus decelerating infall motion of gas into black hole.
This rather promotes radiative cooling, thereby decreasing temperature.
This means, the parameter $\mu$ more appreciably affects 
the flow structure in the vicinity of the central hole.

\subsection{Observable $\mu$ effects}

Finally, we show the effects of changing $\mu$ more explicitly 
in the model spectral fitting diagram in comparison 
with the observational properties of soft X-ray dominated sources. 
The method of model fitting is the same as 
that described in Mineshige et al. (1994; see also Watarai et al. 2000). 
The fitting parameters are
$\Rinm$, which determines the height of the spectral peak, 
$\Tinm$, specifying the frequency of the spectral peak, and
the power-law index in the temperature distribution, 
``$p$''$(\equiv -d\ln T/d\ln r)$,
which controls the spectral slope at low-energy (the Rayleigh-Jeans) side
of the peak.  We thus assume the form,
\begin{equation}
\label{SEDfit}
  S(E) \propto \int_{\Tinm}^{T_{\rm out}^{\rm mod}} 
    \left(\frac{T}{\Tinm}\right)^{-2/p-1}B(E, T) \frac{dT}{\Tinm},
\end{equation}
where $T_{\rm out}^{\rm mod}$ is the temperature at the outer edge 
and $B(E, T)$ is the Planck function. 
The limit of $p \sim 0.75$ corresponds to 
the original MCD model, 
while $p \sim 0.5$ indicates the typical temperature gradient 
of the slim disk model, $T \propto r^{-1/2}$. 
The results of the fitting for different $\mu$ values 
are summarized in Tables 1--3 
(see also Tables 4 and 5 for different $\alpha$ cases).
Throughout this section, we fix $\alpha=0.1$.
Note that we are more concerned with relative changes of the
fitting parameters as functions of $M$ and $\dot M$, 
rather than their absolute values for specific values of $M$ and $\dot M$,
since realistically, general relativistic effects and other effects
(such as self-shielding and the disk tilting) should modify the results,
which are not easy to evaluate precisely.  However, general
tendency does not change as long as the central black hole is non-rotating
(see Manmoto, Mineshige 2001 for the detailed discussion).

\begin{figure}
  \begin{center}
    \FigureFile(80mm,80mm){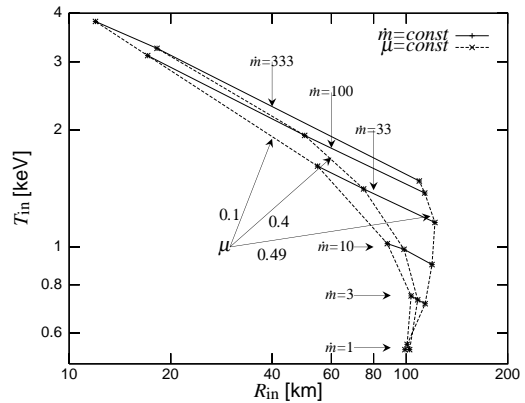} 
  \end{center}
  \caption{
The fitting parameters $\Tin, \Rin$ are plotted in this figure.
The black-hole mass is $M=10M_{\odot}$.
The solid lines are $\dot{m}=const.$
and the dotted lines are $\mu=const.$, respectively.}
\label{fig:RTinfig}
\end{figure}

According to our previous paper (Watarai et al. 2001), 
the ``$\Tin-\Rin$'' diagram is useful for diagnosing the objects 
having a soft thermal component in the X-ray spectrum; 
with this we can constrain various models (see also Belloni 2000, Mizuno et al. 2000). 
We considered the spectral hardening factor, $\xi$ 
(Shimura, Takahara 1995), and the correlation factor, 
$\eta$ (Kubota at al. 1998), in the fitting results. 
The hardening factor $\xi$ represents the spectral hardening due 
to Comptonization processes within the disk 
($\Tin=\xi \Tinm$, see the second column of tables). 
The $\eta$ parameter is introduced, since the radius of the apparent inner edge, $\Rinm$ derived by spectral fitting based on equation (\ref{SEDfit}) 
is generally overestimated by a factor of few 
because of negligence of the boundary term, 
so as to remove such artificial effects and to evaluate the radius of
the real inner edge ($\Rin=\eta \Rinm$, see the fourth column of tables). 
 In this paper, we set ($\xi, \eta)=(1.7, 0.41)$ according to Kubota et
 al. (1998). 
The ``$\Tin-\Rin$'' diagram are shown in Fig \ref{fig:RTinfig}.

In the cases with $\mu=0.01$, the fitting results 
are nearly consistent those with $\mu=0$. 
As $\mu$ increases, $\Rin$ decreases (see Tables 1--3).
For $\mu>0.4$, in particular, 
the inner disk temperature can never be larger than 2.0 keV. 
As we described in the previous section, when the effective 
$\alpha$ is large, the disk cannot release 
its gravitational energy efficiently even for $\dot{m}=100$. 

\begin{figure}
  \begin{center}
    \FigureFile(80mm,80mm){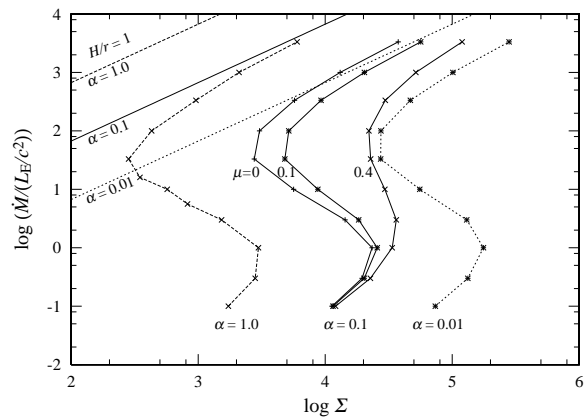} 
  \end{center}
  \caption{The thermal equilibrium (S-shaped) curves for
different parameter sets; ($\alpha$, $\mu$)=(1.0, 0.0), (0.1, 0.0), (0.1,
 0.1), (0.1, 0.4), and (0.01, 0.0), respectively. The abscissa is the normalized
mass accretion rate, and the ordinate is the surface density both on the
 logarithmic scale.
The dashed-dotted lines represent the loci of $H/R=1$ for several value of $\alpha$.
}
\label{s-curve}
\end{figure}

It is difficult to discriminate different models from 
the observational fitting results in Fig \ref{fig:RTinfig}, 
because the difference between $\mu=0$ and 0.4 is only a factor of a few. 
In the case of smaller $\dot{m}$ value, 
there is practically no change. When $\dot{m}$ is large, 
$\dot{m} \gtrsim 30$, on the other hand, 
$\Tin$ changes by factor of a few, 
and $\Rin$ changes by one order of magnitude, depending on $\mu$ values.

\section{Discussion}

We thus expect different time-dependent behavior of a single source,
depending on the $\mu$values, if large $\dot{m}$ modulation is observed.
In the case of $\mu \sim 0.5$, importantly,
the disk inner radius $\Rin$ does not change much even when $\dot{m}$
changes a lot, in contradiction with the observed behavior of IC~342
which clearly shows a decrease in $\Rin$ when $L$ increases
(Watarai et al. 2001).
Thus, the case with $\mu \sim 0.5$ can be ruled out at least for the
 case of ULXs.

According to the SSD theory, the $\alpha$ parameter affects the 
flow velocity ($v_r$) and thus matter density ($\Sigma$),
in such a way that mass-flow rate ($\propto v_r \Sigma$) is kept constant,
but does not largely affect its local emitting flux 
(as long as a simple blackbody radiation is assumed).
In the high accretion-rate systems, however, 
the temperature within the transonic region of the flow
is directly affected by the $\alpha$ value.
Unlike the SSD the thickness of the high-luminosity disk 
becomes mildly thick (the relative thickness $H/R$ 
is less than but of order unity) and Thomson optical depth is quite large.
Therefore, photons generated by viscosity take longer time to reach
the disk surface than the accretion timescale so that
the generated energy by viscosity cannot be radiated immediately. 

From the observational point of view, the ``$\mu$'' parameter acts 
more effectively on the fitting parameters in the slim disk 
than in the standard disk. This result leads the conclusion that 
the super-Eddington sources are useful for investigating the 
nature of viscosity. We find that, even if objects have truly high accretion rates, they may look like a standard-type disk, if $\mu$ is large.
If we accurately measure the change in $\Rin$ 
with changes of $\dot{m}$, we will be able to give a good estimate
to the $\mu$ value.

The origin of the viscosity in accretion disk is still poorly known,
so it is practically impossible at the present to constrain the value of $\mu$ 
from the first principle.
Also, the observational determination of $\mu$ is not easy.
If $\mu$ has a larger value in reality, 
the disk does not enter the advection-dominated regimes,
even if the super-critical accretion is realized.
The results of the present study are useful for constraining
the viscosity model. For example, large $\mu~(>0.1)$ models do not agree 
with the observed properties of ULXs, which show a tendency of decreasing $\Rin~(\lesssim 3r_{\rm g})$ at large luminosity (Watarai et al. 2001).
Hence, the viscosity model with a large $\mu$ parameter 
can be ruled out for such sources.

Figure \ref{s-curve} illustrates the thermal equilibrium curves of 
the calculated disk at a fixed radius ($r=7r_{\rm g}$). 
The straight lines in the upper-left corner of the figure represent the loci of $H/r=1$ 
for several values of $\alpha$ =0.01, 0.1, and 1.0 from the bottom, respectively. 
%
%
How close the equilibrium curve is to the line of $H/r=1$ 
is a good indicator to assess how the advection is important in
energy equation, since 
$Q_{\rm adv} \gtrsim (H/r)^2 Q_{\rm vis}$ (Kato et al. 1998). 
The relative distance between the upper branch of the S-shaped curve and 
the line of $H/r=1$ hardly changes for each set of the same $\alpha$ 
in Fig \ref{s-curve}.
On the other hand, it is clearly demonstrated 
 that when $\mu$ is large, the kinks are not strongly manifested. 
Limit-cycle oscillations can occur at high
$\dot m$, only when $\mu$ is small, and the amplitudes of
oscillations decreases as $\mu$ increases
(Honma et al. 1991; Szuszkiewicz, Miller 1997, 1998). 
Large $\mu$ values tend to stabilize the disk structure 
so that for even higher $\mu (\gtrsim 0.5)$ the limit cycle oscillations cannot occur. 

The occurrence of limit-cycle oscillations 
in real systems is still in open question. 
There is a hint of such oscillation in GRS1915+105 (Yamaoka et
al. 2001), but we need careful data analysis in future work.

The complementary work to the present study is made by
Manmoto, Mineshige (2001) who consider fully relativistic
treatments of the slim-disk problem but only for the canonical
viscosity prescription ($t_{r\varphi} = -\alpha p_{\rm tot}$) with
$\alpha = 0.1$.  Main results are:
(i) for the disk around a Schwarzschild black hole
$\Rin$ $decreases$ from 3$\rS$ to $\sim 1.5 \rS$ 
as mass-accretion rate increases beyond a critical value 
${\dot M}_{\rm crit} \sim 30 L_{\rm E}/c^2$;
(ii) for the disk around a Kerr hole (KBH), conversely,
$\Rin$ $increases$ from $< \rS$ to $\sim 1.5 \rS$ 
as mass-accretion rate increases because of significant
self-shielding of the innermost hot region;
(iii) at accretion rate above the critical value, 
disks around Schwarzschild and Kerr holes
look quite similar;
(iv) in both cases, the temperature distribution is
somewhat flatter, $\propto r^{-1/2}$, at large luminosity.
These results are basically in good agreement with the present
study based on the pseudo-Newtonian treatment (except
for the case of Kerr holes which are not considered here),
but there exist some quantitative differences.  In the present study,
hence, we are more concerned with the qualitative trend
and thus it is unlikely that our main conclusions significantly change,
even if fully relativistic treatments expect that very small $\Rin <
\rg$ may be an artifact of our pseudo-Newtonian treatments.

For more realistic study along this line we also need 
fully 2D studies, taking into account the radiative transfer 
towards the vertical direction and various MHD processes.
This is left as future work.
Time dependent modeling of bursting behavior observation GRS1915+105 is
another interesting subject in future work.

Recently, King et al. (2001) proposed that large $L$ and high $\Tin$
observed in ULXs could be understand, if X-ray is moderately beamed. We
suggest that spectral fitting with the $p-free~model$ will be able to test
this hypothesis, since it only the thermal emission from the innermost
region is beamed, the resultant spectrum will be of a single-temperature
blackbody; i.e., large $p$ will be obtained. 
Otherwise, we expect $0.5 \le p \le 0.75$.
We can thus judge it beaming is substantial in future observations (with,
e.g., XMM Newton).
\section{Conclusions}

We have discussed the properties of the slim disk
for different viscosity prescriptions and their soft X-ray signatures. 
Our conclusions are summarized as below.
\begin{enumerate}
\item Some $\alpha$ dependence is found 
in the radial temperature distribution of the innermost region 
of high accretion-rate systems ($\dot{m} \gtrsim 30$);
an increase in $\alpha$ tends to increase
 the temperature of the transonic flow, 
making a spectrum somewhat harder, although the change
is not large and thus not easy to detect observationally.

\item When the parameter $\mu$ takes a larger value 
(e.g. $\mu=0.5$), the slim-disk nature  (small $\Rin \lesssim 3r_{\rm g}$ and flatter temperature profile, $T_{\rm eff} \propto r^{-1/2}$) is not appreciable even
at high luminosity comparable to the Eddington luminosity.

\item The origin of the viscosity is still unknown, but observations of high luminosity objects will be able to discriminate 
different viscosity prescriptions for accretion disks. 
For example, large $\mu$ values are not incompatible with the
behavior of some ULXs.  The viscosity should thus be more dependent
on total pressure, rather than gas pressure solely.
\end{enumerate}


We are grateful to the referee, E. Szuszkiewicz, for pointing out the
initial of errors
in the first version and for useful comments and discussion, which
helped our making the revised version. 
We would also like to thank M. Takeuchi and T. Kawaguchi for discussion.
This work was supported in part by the Grants-in Aid of the Ministry of
Education, Science, Sports, and Culture of Japan (13640238, SM).

\begin{longtable}{rrrrr}
  \caption{Results of fitting with $p$-free model (0.2-10keV).
($M=10M_\odot$, $\mu=0.1$, $\alpha=0.1$)
}
  \label{tab:LTsample}
  \hline\hline
  $\dot{M}/(L_{\rm E}/c^2)$ & 
  $\Tin$ (keV) & 
  $\Rinm$ (km) &
  $\Rin/r_{\rm g}$ &
   ~$p$~~ \\
  \hline
  \endfirsthead
  \hline
  \endlastfoot
   1  & 0.56 \footnotesize {(0.2-8.0 keV)} & 245.4 & 3.35 & 0.75 \\ 
   3  & 0.75   & 237.1 & 3.24 & 0.74 \\ 
  10  & 1.02   & 211.0 & 2.88 & 0.72 \\ 
  33  & 1.60   & 142.1 & 1.78 & 0.68 \\ 
 100  & 2.86   &  56.7 & 0.77 & 0.61 \\ 
 333  & 3.37   &  40.1 & 0.55 & 0.58 \\ 
\end{longtable} 


\begin{longtable}{rrrrr}
  \caption{
Results of fitting with $p$-free model ($M=10M_\odot$, $\mu=0.4$, $\alpha=0.1$)
}
  \label{tab:LTsample}
  \hline\hline
  $\dot{M}/(L_{\rm E}/c^2)$ & 
  $\Tin$ (keV)  & 
  $\Rinm$ (km) &
  $\Rin/\rg$ &
   ~$p$~~ \\
  \hline
  \endfirsthead
  \hline
  \endlastfoot
1    & 0.54~\footnotesize{(0.2-5.0keV)}  & 249.6 & 3.41 & 0.75 \\ 
3    & 0.75   & 247.0 & 3.37 & 0.74 \\ 
10   & 0.99   & 233.5 & 3.19 & 0.73 \\ 
33   & 1.39   & 200.1 & 2.73 & 0.71 \\ 
100  & 1.90   & 138.1 & 1.89 & 0.65 \\ 
333  & 2.92   &  63.8 & 0.87 & 0.59 \\ 
\end{longtable}

\begin{longtable}{rrrrr}
  \caption{
Results of fitting with $p$-free model ($M=10M_\odot$, $\mu=0.49$, $\alpha=0.1$)
}
  \label{tab:LTsample}
  \hline\hline
  $\dot{M}/(L_{\rm E}/c^2)$ & 
  $\Tin$ (keV)    & 
  $\Rinm$ (km) &
  $\Rin/\rg$ &
   ~$p$~~ \\
  \hline
  \endfirsthead
  \hline
  \endlastfoot
1    & 0.54~\footnotesize{(0.2-8.0keV)} & 241.77 & 3.36  & 0.74   \\ 
3    & 0.73  &  266.24 & 3.69  & 0.76   \\ 
10   & 0.90  &  290.92 & 4.04  & 0.75   \\ 
33   & 1.16  &  296.76 & 4.10 & 0.73   \\ 
100  & 1.38  &  276.28 & 3.83  & 0.68   \\ 
333  & 1.48  &  266.66 & 3.07  & 0.62   \\ 
\end{longtable}

\begin{longtable}{rrrrr}
  \caption{
Results of fitting with $p$-free model ($M=10M_\odot$, $\mu=0$, $\alpha=0.01$)
}
  \label{tab:LTsample}
  \hline\hline
  $\dot{M}/(L_{\rm E}/c^2)$ & 
  $\Tin$ (keV)    & 
  $\Rinm$ (km) &
  $\Rin/r_{\rm g}$ &
   ~$p$~~ \\
  \hline
  \endfirsthead
  \hline
  \endlastfoot
1    &  0.56~\footnotesize{(0.2-8.0keV)}  &  245.4      & 3.41  & 0.75   \\
3    &  0.73  &  236.4      & 3.28  & 0.74   \\
10   &  1.04  &  207.0      & 2.87  & 0.71   \\
33   &  1.51  &  153.5      & 2.13  & 0.68   \\
100  &  2.50  &  68.64      & 0.95  & 0.61   \\
333  &  2.94  &  51.59      & 0.72  & 0.57   \\
\end{longtable}

\begin{longtable}{rrrrr}
  \caption{
Results of fitting with $p$-free model ($M=10M_\odot$, $\mu=0$, $\alpha=1.0$)
}
  \label{tab:LTsample}
  \hline\hline
  $\dot{M}/(L_{\rm E}/c^2)$ & 
  $\Tin$ (keV)  & 
  $\Rinm$ (km) &
  $\Rin/\rg$ &
   ~$p$~~ \\
  \hline
  \endfirsthead
  \hline
  \endlastfoot
1    &  0.56~\footnotesize{(0.2-8.0keV)} & 244.2 & 3.39  & 0.75   \\
3    &  0.75  & 234.9     & 3.26  & 0.74   \\
10   &  1.04  & 204.0     & 2.83  & 0.71   \\
33   &  3.09  & 46.04     & 0.64  & 0.65   \\
100  &  3.93  & 36.82     & 0.51  & 0.62   \\
333  &  4.18  & 30.67     & 0.43  & 0.58   \\
\end{longtable}




\end{document}